\documentclass[reprint, aps, pre, a4paper, floatfix]{revtex4-1}
\usepackage{graphicx}
\usepackage[english]{babel}
\usepackage[utf8]{inputenc}
\usepackage{url} 
\usepackage{amsmath}
\usepackage{float}
\usepackage{verbatim}
\usepackage{color}
\renewenvironment{description}[1][0pt]
 {\list{}{\labelwidth=0pt \leftmargin=#1
  }}
 {\endlist}

\begin{document}

\title{Master equation for the degree distribution of a Duplication and Divergence network}
\author{Vítor Sudbrack}
\author{Leonardo G. Brunnet}
\author{Rita M. C. de Almeida}
\author{Ricardo M. Ferreira}
\author{Daniel Gamermann}
\affiliation{Instituto de Física - Universidade Federal do Rio Grande do Sul (UFRGS). Av. Bento Gonçalves, 9500}

\begin{abstract}
Network growth as described by the Duplication-Divergence model proposes a simple general idea for the evolution dynamics of natural networks. In particular it is an alternative to the well known Barab\'asi-Albert model when applied to protein-protein interaction networks.
In this work we derive a master equation for the node degree distribution of networks growing via Duplication and Divergence and we obtain an expression for the total number of links and for the degree distribution as a function of the number of nodes. Using algebra tools we investigate the degree distribution asymptotic behavior. 
Analytic results show that the network nodes average degree converges if the total mutation rate is greater than 0.5 and diverges otherwise. Treating original and duplicated node mutation rates as independent parameters has no effect on this result. However, difference in these parameters results in a slower rate of convergence and in different degree distributions. The more different these parameters are, the denser the tail of the distribution. 
We compare the solutions obtained with simulated networks. These results are in good agreement with the expected values from the derived expressions. The method developed is a robust tool to investigate other models for network growing dynamics.
\end{abstract}

\maketitle

\section{Introduction}

Physicists are required to build models that extract the essence of observable phenomena seen in nature in order to understand and describe them. As physics endeavors in studying complex phenomena in distinct fields such as biology or social sciences, a commonly applied paradigm is the use of graph theory. In this approach the system under study is described as a network consistent of a set of nodes and a set of links among them.

Examples of systems studied within this approach are social networks \cite{socialnet}, author citations \cite{citenet}, flights connections \cite{flightnet}, metabolic models \cite{metnet1,metnet2}, protein-protein interactions \cite{protnet1,protnet2}, electrical grids \cite{gridnet1,gridnet2}, among others.

In the present work we are interested in \textit{Network dynamics}, i.e. the study of the dynamics behind a growing graph. The motivation lies in the dynamics of biological networks, given that many such systems find a natural description within those models. Understanding the evolutionary selection rules resulting in networks with similar topological characteristics as the real observed ones, may give insights about the underlying biological processes (natural selection) behind these structures. For example, the protein association networks can help understand the evolution of species' genomes \cite{ohno, wen, ispolatov, taylor}.

In the present work, the network dynamics is described as a Markovian process. Within this approach, the network state at a given time depends only on its configuration on the previous moment. Given a set of rules that describe how the network changes in each time step, we construct the corresponding master equation representing the evolution of the system's configurations. A similar methodology was used in the work by Ferreira {\it et al.} \cite{Ferreira_2016}. With this approach, they presented analytical results and simulations of networks growing according to the Barábasi-Albert rule. Here we focus on a complementary approach to model protein-protein network dynamics \cite{Ferreira_2003}  and explore the evolution of an adapted version  of the Duplication Divergence model \cite{Vazquez_2003}. The importance in describing the average behavior of stochastic processes in this manner is to know the network behavior for different values of the parameters without the need of long, time consuming, numerical simulations to obtain statistically relevant information. 

This article is organized as follows. In the next section we focus in explaining the Duplication and Divergence model. Following the model explanation, we derive an expression for the total number of links as a function of the number of nodes, which gives us a straight forward way to obtain the mean degree of the graph. Then we study the graph growth as a Markovian chain, in which the next degree distribution of the network is a function of the current degree distribution, pondered by the probabilities of all possible occurrences in each time step. Finally, we study the asymptotic limit of the degree distribution.


\section{The Model}

Given an initial small network (three nodes connected to each other forming a triangle) we study the Markov process where, in each time step a node of the network is randomly chosen to be copied i.e. a new node is created with exactly the same neighbors as the chosen one. In what follows we refer to the copied node as original and its copy as duplicated. 
After duplication, original and duplicated nodes may diverge, meaning that each link of the original node is lost with probability $m_o$ and each link in duplicated node is lost with probability $m_d$.
Also, a link between original and the duplicated nodes is always added.

This model is an adaptation of the Duplication and Divergence model, originally developed by Vázquez et. al. \cite{Vazquez_2003, modeling}. In the model proposed by Vázquez a new node is also added by copying an existing node and all its links. New node and ancestor are linked with a probability $p$. Also, either the link between new node and a third neighbor or the link between ancestor and this neighbor is lost  with probability $q$ \cite{modeling}. Our model treats independently the loss of a link by original and duplicated nodes and sets $p = 1$.

In V\'azquez's model the network growth is based on local rules, that is, rules that require only information on one node instead of rules that require information over all the network. This model of network dynamics applied to protein-protein interaction networks allows all proteins to evolve from a common ancestor through gene copies (represented by duplications) and mutations (divergence). Therefore, it would mimic the entire history of a genome evolution \cite{modeling}. 

\section{Mean degree and total number of links}

First, let's evaluate the behavior of the network average node degree $\bar{k}$ as a function of the number of nodes in the network, $t$. 
Given that a node with degree $k$ is chosen to be duplicated, the number of links in the next step changes. The mean change in the number of links is given by\footnote{The bar over a term in the equation indicates that the average of the process will be considered.}:

\begin{eqnarray}
L_{t+1} &= L_{t} + \overline{(k - m_ok - m_dk + 1)} \nonumber \\
&= L_{t} + (1 - m_o - m_d)\bar{k} + 1
\label{L_discreta}
\end{eqnarray}

Terms in the right-hand side of Eq. (\ref{L_discreta}) represent, from left to right: existing links, links added due to the duplication of a $k$-degree node, mean number of lost links of the original node, mean number of lost links for the duplicated node and creation of the original-duplicated link.

Let's define the total mutation parameter $M$ as the sum of the independent mutation parameters $m_o$ and $m_d$: $M=m_o+m_d$. Now, given the relation between node's degree distribution and number of links in a network, known as the handshake lemma \cite{euler},
\begin{equation}
2L = \sum_{i=1}^N k_i = N\bar{k},
\end{equation}
and, considering the time as the number of nodes (since we add a node in each time step), the following map for the network average degree is written

\begin{equation}
\frac{(t+1)\bar{k}_{t+1}}{2} = \frac{t\bar{k}_{t}}{2} + (1-M)\bar{k}_{t} + 1\;.
\end{equation}

This equation can be rearranged to explicit the $\bar{k}_{t+1}$ term:

$$
\bar{k}_{t+1} = \frac{(t+2-2M)\bar{k}_{t}+2}{t+1}
$$

For long times (big values of $t$), this can be approximated as a continuous differential equation:

$$
\frac{\partial \bar{k}}{\partial t} = \frac{1-2M}{t+1}\bar{k} + \frac{2}{t+1}
$$

For $0 \leq M \leq 2$, the solution of this ODE is:

\begin{equation}
 \bar{k}(t) =
 \begin{cases}
c(t+1)^{1-2M} + \frac{2}{2M-1} &\quad\text{for\ }M\neq 0.5;\\
2\log(t+1) + c &\quad\text{for\ }M= 0.5\;,
\end{cases}
\label{mean_degree}
\end{equation}
where the constant $c$ is related to network initial conditions.

For instance, consider the case of a process without divergence ($M=0$). Starting with a triangle ($\bar{k}(3)=2$), the duplication will make every node connected with every other, i.e., a complete graph. In this particular case Eq. (\ref{mean_degree}) results in
$$
\bar{k}(t) = (t+1)-2 = t-1.
$$

In the limit $t \to \infty$, the possible asymptotic behaviors for Eq. (\ref{mean_degree}) are:

\begin{description}[1cm]
 \item[For M $>$ 0.5:] The mean degree converges to $\frac{2}{2M-1}$.
 \item[For M $<$ 0.5:] The mean degree diverges as $t^{1-2M}$.
 \item[For M $=$ 0.5:] The mean degree diverges logarithmically.
\end{description}

\begin{figure}
	\begin{center}
	\includegraphics[width=\columnwidth]{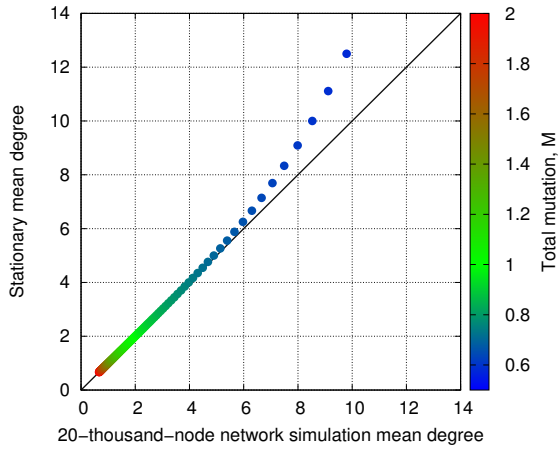}
	\caption{[color on-line] Scatter plot of the average degree for 20 thousand node networks obtained numerically compared with the result given by Eq. (\ref{mean_degree}) as a function of the total mutation $M=m_o+m_d$. Note that the greater the total mutation $M$, the faster the stationary limit is reached.}
	\label{scatterplot_graumed}
	\end{center}
\end{figure}

We represent these possible situations in Fig. \ref{scatterplot_graumed}, where each point refers to a network evolved through the model dynamics with a different value for the total mutation parameter $M$. In the vertical axis one has the correspondent mean degree calculated by Eq. (\ref{mean_degree}) (in the limit $t \to \infty$) and in the horizontal axis it is presented the results for the mean degree evaluated from simulated networks growing according to the model rules. The correspondence of the points colors with the parameter $M$ can be read in the color scale. The figure clearly shows that, for high values of the total mutation $M$ (values equal or bigger than $0.8$), a network with 20 thousand nodes has already reached its stationary behavior (the term $t^{1-2M}$ can be neglected). The closer the total mutation gets to $M=0.5$, the slower one observes the convergence to the stationary limit. For values of total mutation under $0.5$ it is expected that the mean degree diverges and so are the points in this figure departing from the main diagonal.

Using again the handshake lemma in equation (\ref{mean_degree}), we obtain the total number of links as a function of time

\begin{equation}
 L(t) = 
 \begin{cases}
ct(t+1)^{(1-2M)} + \frac{t}{2M-1} &\quad\text{for\ }M\neq 0.5;\\
t\log(t+1) + ct &\quad\text{for\ }M= 0.5
\end{cases}
  \label{links_Eq}
\end{equation}

\begin{figure}
	\begin{center}
	\includegraphics[width=\columnwidth]{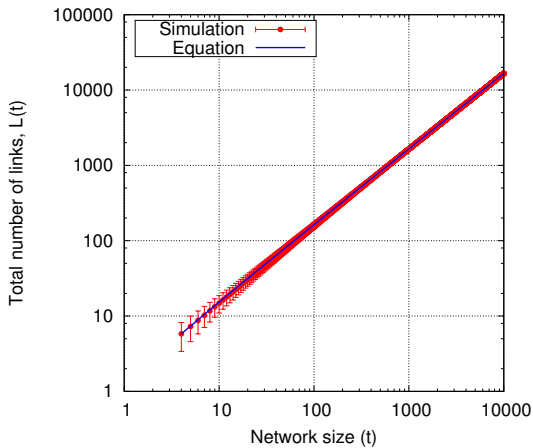}
	\caption{[color on-line] A comparative between the results of the average (over 5 thousand networks) total number of links as a function of the network size and the result given by Eq. (\ref{links_Eq}) for $M=m_o+m_d=0.8$. }
	\label{links_graph}
	\end{center}
\end{figure}

In Fig. \ref{links_graph} we show results comparing the mean value for the number of links in 5 thousand networks as a function of time with the expected value found through from Eq. (\ref{links_Eq}). 

\section{Degree distribution for finite times}
	
Considering the model dynamics, it is possible to establish relations among the number of nodes with degree $k$ at time $t$, $N(k,t)$, and at time $t+1$:
 
\begin{align}
N(k,t+1) &=& N(k,t) \nonumber \\
&-& \frac{N(k,t)}{t} \nonumber \\
&+& \sum_{i=k-1}^{t-1}P_o(i \to k)\frac{N(i,t)}{t} \nonumber \\
&+& \sum_{i=k-1}^{t-1}P_d(i \to k)\frac{N(i,t)}{t} \nonumber \\
&+& \frac{(k-1)N(k-1,t)}{t}(1-m_o)(1-m_d) \nonumber \\
&+& \frac{(k+1)N(k+1,t)}{t}m_om_d \nonumber \\
&-& \frac{kN(k,t)}{t}[(1-m_o)(1-m_d) + m_om_d],  \label{eq_discreta_N_k}
\end{align}
the terms in this equation represent, from top to bottom: the number of existing nodes of degree $k$; the probability of a $k$-degree node to be chosen to duplicate, leaving this degree; the first sum is the total probability that a node of degree different from $k$ is duplicated and after the divergence process becomes a $k$-degree node; the second sum is the same total probability for the duplicated node. Finally, the last three terms are the probabilities that the neighbors of the node chosen to be duplicated arrive, from a different degree (either $k+1$ or $k-1$) to the degree $k$ and the probability that a $k$-degree neighbor goes to $k+1$ or $k-1$. 

Since each link, during the divergence process, is lost independently with probability $m_o$ or $m_d$, the probabilities inside the sums are binomial distributions representing the probabilities that either the original or duplicated node goes from degree $i$ to degree $k-1$ after the divergence, and finally received the original-duplicated link:

\begin{subequations}
\begin{eqnarray}
P_o(i \to k ) = C^{i}_{k-1}(1-m_o)^{k-1}(m_o)^{i-k+1} 
\label{prob_i_to_k1}
\\
P_d(i \to k ) = C^{i}_{k-1}(1-m_d)^{k-1}(m_d)^{i-k+1} 
\label{prob_i_to_k2}
\end{eqnarray}
\end{subequations}

It is important to note the domain of the above functions (\ref{prob_i_to_k1}) and (\ref{prob_i_to_k2}), $i\geq k-1$, that is, the node after passing through duplication and divergence can increase its degree by one unit (which means keeping all its links and adding the copied-duplicated link), keep the same degree (losing one neighbor and adding the copied-duplicated link) or else it will have its degree decreased due to the loss of more than one neighbor.

One can verify the addition of a single node in each time step by summing the master equation for $N(k,t)$, Eq. (\ref{eq_discreta_N_k}), over all possible degrees

\begin{equation}
\sum^{\infty}_{k=0}(N(k,t+1)-N(k,t)) = 1.
\label{condicao}
\end{equation}

\begin{figure}
	\begin{center}
	\includegraphics[width=\columnwidth]{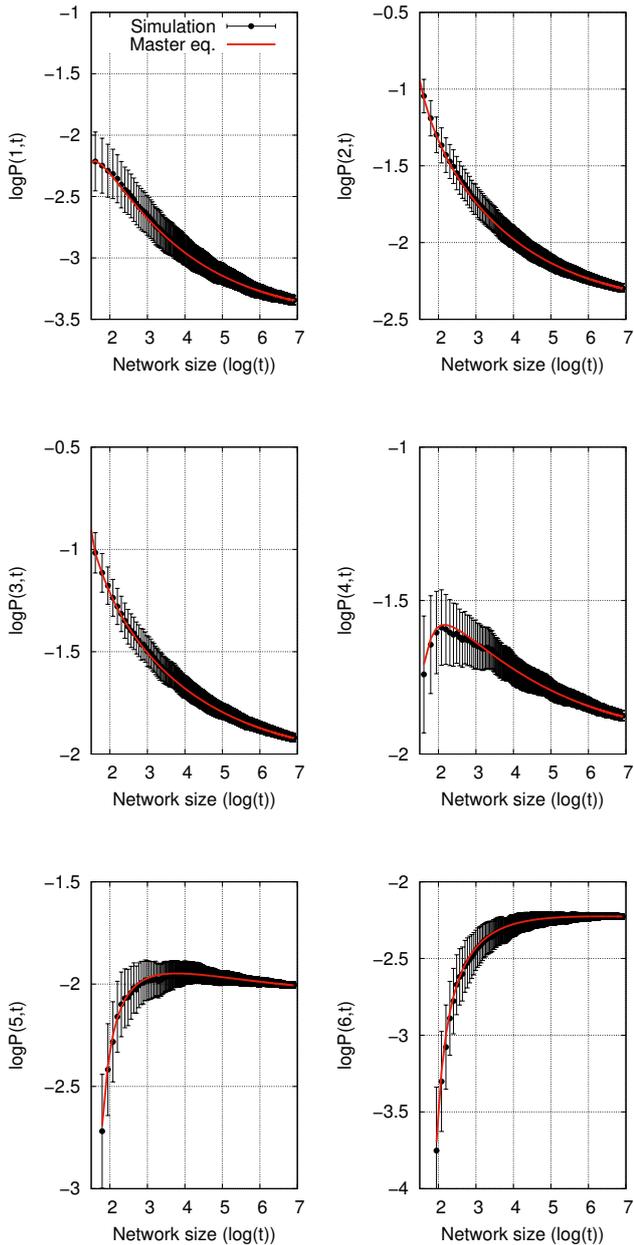}
	\caption{[color on-line] Comparison of the results for the average fraction ($P(k,t)=N(k,t)/t$) of nodes of degree $k$ in $5$-thousand networks simulated from $3$ nodes until $10^{3}$ with the results predicted by the numerical integration of equation (\ref{eq_discreta_N_k}) for $m_o=0.25$ and $m_d=0.40$ (resulting in $M=0.65$).}
	\label{multiplot}
	\end{center}
\end{figure}

As shown in Fig. \ref{multiplot}, the numerical solution of the map in equation (\ref{eq_discreta_N_k}) produces results in excellent agreement with the mean number of nodes with degree $k$ at the time $t$ evaluated from thousands of simulated networks growing through the stochastic process.

As an analytic example, using $m_o=0$ e $m_d=0$, no links are lost by neither node, and equation (\ref{eq_discreta_N_k}) becomes,

\begin{align}
N(k, t+1) &=& N(k,t) - \frac{N(k,t)}{t} + 2\frac{N(k-1,t)}{t} \nonumber \\
&+& \frac{(k-1)N(k-1,t)}{t} - k\frac{N(k,t)}{t}
\label{above}
\end{align}

Eq. (\ref{above}) can be written as

\begin{equation}
N(k, t+1) - N(k,t) = \frac{k+1}{t}\Big( N(k-1,t) - N(k,t)\Big)
\end{equation}
Note that, in the case where the initial network is totally connected, that is, $k = t-1$, there is only the flux of all nodes having its degree increased by one in a totally connected network, $N(k,t)= t\delta_{k-1,t}$.

\section{Asymptotic Degree distribution}

Equation (\ref{eq_discreta_N_k}) can be conveniently written in its matrix form, defining the column vector $\vec{N}(t)$ whose components are $N_k(t) = N(k,t)$:

\begin{equation}
\frac{d}{dt}\vec{N} = \frac{1}{t} A\vec{N},  
\label{vec_eq}
\end{equation}
  
where the matrix $A$ shows the following elements, constant with respect to time:

\begin{align}
a_{i,k} &=& -(k[(1-m_o)(1-m_d) + m_om_d] + 1)\delta_{k,i} \nonumber \\
&+& (k-1)(1-m_o)(1-m_d)\delta_{k-1,i} \nonumber \\
&+& (k+1)m_om_d\delta_{k+1,i} \nonumber \\
&+& P_o(i \to k)
+ P_d(i \to k)
\label{coef_matriz}
\end{align}

The coefficients of equation (\ref{coef_matriz}) explicit the couplings in equation (\ref{vec_eq}). It is possible to solve equation (\ref{vec_eq}) through a matrix decomposition of the matrix $A_{n\times n}$ in its eigenvalues and eigenvectors: Writing $A=X^{-1}DX$, putting this decomposition in equation (\ref{vec_eq}) and multiplying from the left by the matrix $X$, one decouples the equations and is able to solve it for each component of the vector $\vec{N}$ in the space where the matrix $A$ is diagonal. Returning to the original space one has:

\begin{equation}
\vec{N} = \sum_{i=1}^{n}c_i\vec{X}_it^{\lambda_i}    \label{solution_for_N}
\end{equation}
Where the $\lambda_i$ are the eigenvalues of matrix $A$, $\vec{X_i}$ the respective eigenvectors (the columns of the matrix $X$, properly normalized, are the vectors $\vec{X_i}$) and the constants $c_i$ depend on the initial conditions.

Normalizing the vector $\vec{N}(t)$ one works with the fraction of nodes with degree $k$:

\begin{equation}
\vec{P}(t) = \frac{1}{t}\vec{N(t)},
\end{equation}
note that $P_k(t)$ is the fraction of nodes of degree $k$ at time $t$. The solution (\ref{solution_for_N}) can be written in terms of the new vector as follows:

\begin{equation}
\vec{P} = \sum_{i=1}^{t}c_i\vec{X}_it^{\lambda_i-1}
\end{equation}

In the stationary state, condition (\ref{condicao}) implies that $\lambda = 1$ is an eigenvalue of matrix $A$ (whose left-eigenvector is $(1,1,1,1...,1)$) \cite{MacCluer2000}, and, to conserve  probability, this must be the greatest eigenvalue, and the normalized right-eigenvector which corresponds to the unitary eigenvalue (called main eigenvetor) will be the asymptotic solution of the degree distribution of a network whose growth is governed by the Duplication and Divergence model with mutation rates $m_o$ and $m_d$.

\begin{figure}
	\begin{center}
	\includegraphics[width=0.8\columnwidth]{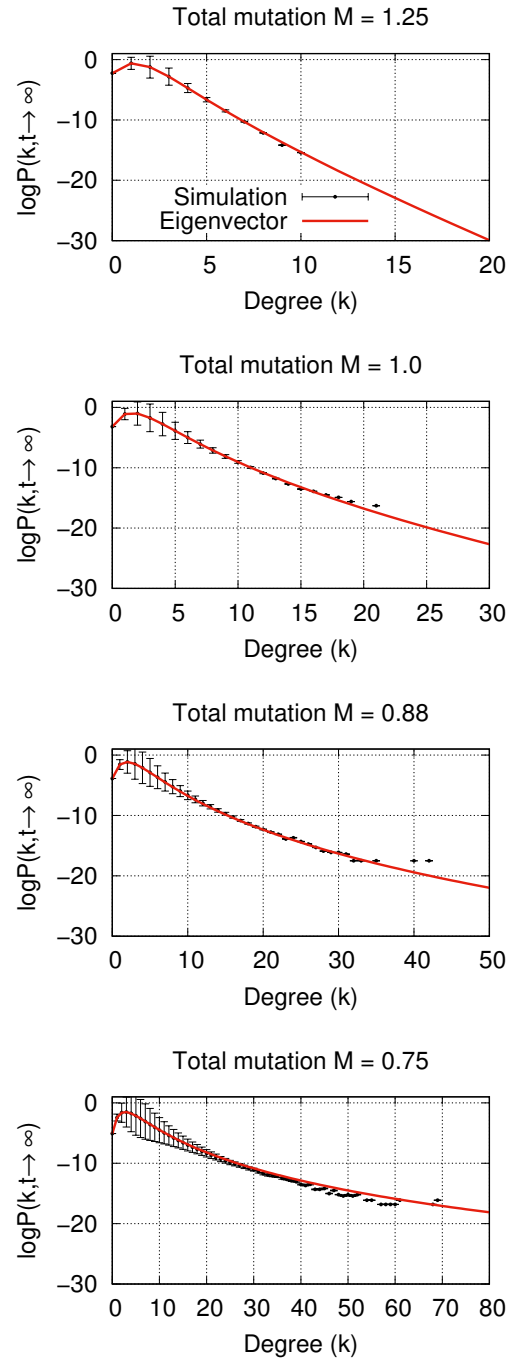}
	\caption{[color on-line] Results of the degree distribution for $t \to \infty$ obtained through the eigenvector decomposition compared to results obtained from simulated networks up to 20 thousand nodes for values of total mutation rates $1.25$, $1.00$, $0.88$ and $0.75$ ($m_o=m_d=M/2$). It is possible to observe that 20-thousand-node networks are large enough to generate the asymptotic degree distribution.}
	\label{autovetores}
	\end{center}
\end{figure}

Fig. \ref{autovetores} compares the main eigenvector numerically obtained with the simulation of networks up to 80 thousand nodes.

Note that the matrix $A_{n \times n}$ has to be truncated for taking the limit $t \to \infty$ because its actual size is $t \times t$ in Eq. (\ref{vec_eq}), which is the greatest degree possible for $(t+1)$-node networks. For $M>0.5$, this truncation still results in a matrix with an eigenvalue whose value is $1$, which is related, as mentioned, to the stationary distribution. When $M<0.5$, any truncation results in a matrix $A$ which has all eigenvalues less than $1$. So when the mean degree diverges, there is obviously no stationary degree distribution.


\section{Parameters $m_o$ and $m_d$ are independent}

As seen in Eq.(\ref{mean_degree}), the mean degree (stationary or time-dependent) is a function that depends only on the total mutation rate, $M=m_o+m_d$. However, Eq.(\ref{eq_discreta_N_k}) cannot be written without considering only the sum of the individual mutations. Therefore, different partitions of the same total mutation lead to different distributions with the same mean. The parameters are interchangeable in both equations which reflects the fact that, after the duplication, the original and copied nodes are indistinguishable.

\begin{figure}
	\begin{center}
	\includegraphics[width=\columnwidth]{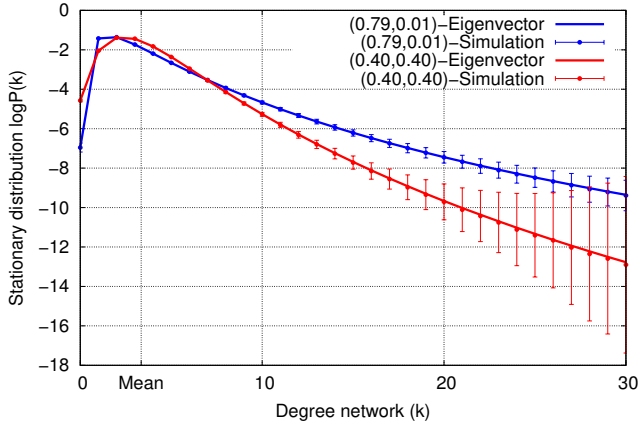}
	\caption{[color on-line] Stationary degree distribution, $P(k, t \to \infty$), for networks with the same total mutation rate, but different mutation rates $m_o$ and $m_d$.}
	\label{comp_est}
	\end{center}
\end{figure}

Finding the stationary distribution numerically for $M=0.80$ in two possible scenarios, shown in Fig.\ref{comp_est}, gives us insights about the network behavior in both cases.

\begin{figure}
	\begin{center}
	\includegraphics[width=\columnwidth]{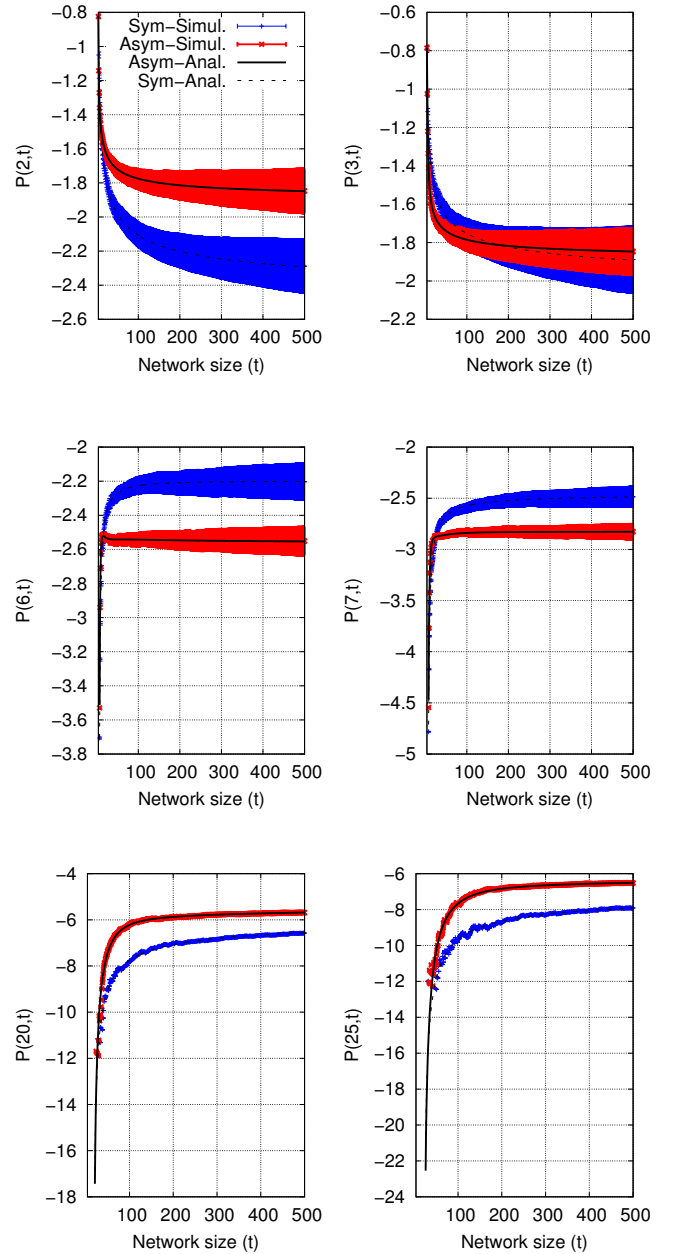}
	\caption{[color on-line] Degree distribution $P(k, t$) as a function of time, for networks with the same total mutation, but different mutation rates $m_o$ and $m_d$. The top graphs are for low degree nodes, the middle ones are near the stationary average and the bottom ones are high degree nodes.}
	\label{comp_est_time}
	\end{center}
\end{figure}

When the mutation parameters are equal ($m_o=m_d=0.4$), one finds a distribution denser near the distribution's average. When the mutation parameters are dissimilar ($m_o=0.79; m_d=0.01$), the resulting distribution has a denser tail, indicating more nodes with lower and higher degrees than the average value. This is a very reasonable result considering the divergence process. In Fig. \ref{comp_est_time}, we show this difference. It is possible to see the time evolution of the degree distribution $P(k,t)$ for degrees in these three cases: lower, near and greater than the stationary average degree (that is $\bar{k}=20/3$, for $M=0.65$) obtained numerically by integrating Eq. (\ref{eq_discreta_N_k}) and simulations.

\begin{figure}
	\begin{center}
	\includegraphics[width=\columnwidth]{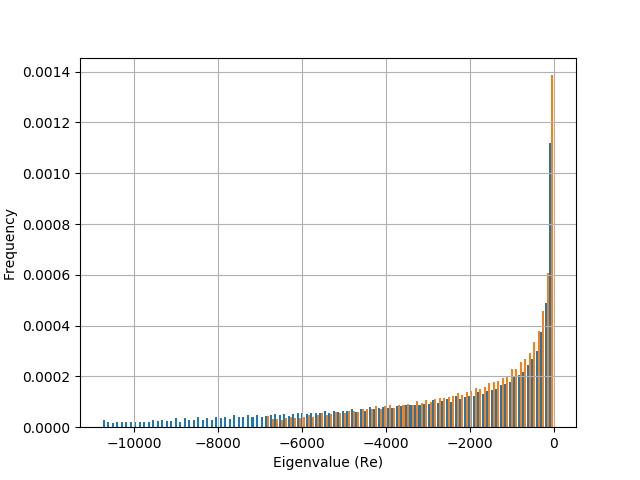}
	\caption{[color on-line] Spectrum of the matrix A for the dynamics with the same total mutation, and $m_o=m_d=0.35$ in blue and $m_o=0.64, m_d=0.01$ in orange.}
	\label{spec}
	\end{center}
\end{figure}

\begin{figure}
	\begin{center}
	\includegraphics[width=\columnwidth]{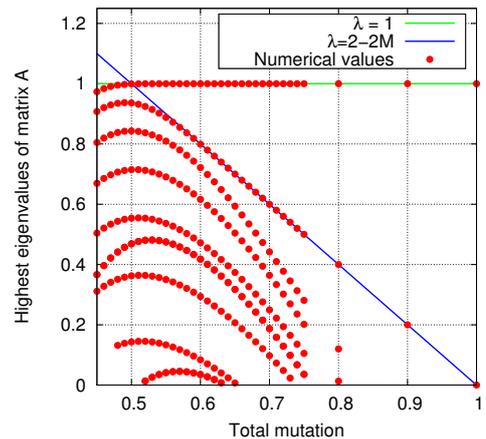}
	\caption{[color on-line] Highest eigenvalues of a size $2000 \times 2000$ matrix $A$ calculated numerically with for $m_o=m_d=M/2$. }
	\label{autovalores}
	\end{center}
\end{figure}

One can also observe the real part of the eigenvalues of matrix $A$ in both cases, shown in Fig. \ref{spec}. The imaginary part is symmetric, which only reflects the fact that the elements of the matrix are real, and therefore the complex eigenvalues come in pairs as complex conjugates. From the real part of the eigenvalues one can infer the convergence rate. Therefore, one can conclude that for different values of $m_o$ and $m_d$ the convergence will take longer, since the eigenvalues are greater. In the symmetrical case, there are smaller eigenvalues than in the asymmetrical case.
The rate of convergence to the stationary solution can also be inferred by the second highest eigenvalue of matrix $A$. In Fig.~\ref{autovalores}  dots represent the real part of the highest eigenvalues of $A$. It is possible to see that the second highest eigenvalue of matrix $A$ for total mutation rate higher than $0.55$ will be given by $2-2M$, and therefore $\vec{P}(t) \propto t^{1-2M}$, which is the same exponent of the convergence of the mean degree of Eq. (~\ref{mean_degree}).

\section{Conclusions}

In the present work, we derived the master equation for the degree distribution of networks evolving through the duplication and divergence model considering the mutation rates of original and duplicated nodes as independent parameters. Numerical integration of the resulting maps agree well with the average values obtained from simulated networks.

The maps obtained for the network nodes average degree and number of links can be approximated as ordinary differential equations and solved analytically. The asymptotic solution for these ODEs, agrees well with simulated data for values of total mutation greater than $0.5$, limit for which the network converges to a stationary degree distribution.

Though the average node degree of a network evolving through this process only depends on the total mutation rate parameter, its degree distribution and rate of convergence will depend on the mutation parameter difference between original and duplicated nodes. The behavior of this distribution can be studied from the spectrum of the matrix $A$, which contains the transition probabilities between nodes of different degrees in each time step of the Markovian process. The more similar the two parameters are, the faster the distribution converges because, in this case, one has less eigenvalues close to 1, and the less dense the distribution tail will be.

To use a Markovian process in order to describe network evolution, allows one to obtain iterative maps for the time evolution of the graph's properties in a wide variety of models. In the asymptotic limit, the maps can be approximated as ODEs, and therefore solved such that one is able to study qualitatively the dynamics dependence on the model parameters without the need to run computationally intensive numerical simulations. Finally, different network growing models may be treated using the present method. In particular, a work considering  a hybrid Barabasi-Duplication/Divergence model is under development.

\section{Acknowledgments}

We thank the contributions of collaborators from MTC-IF-UFRGS. We also thank CNPq for the financial support.

\bibliographystyle{ieeetr}
\bibliography{paper}

\begin{thebibliography}{10}

\bibitem{socialnet}
S.~P. Borgatti, A.~Mehra, D.~J. Brass, and G.~Labianca, ``Network analysis in
  the social sciences,'' {\em Science}, vol.~323, pp.~892--895, feb 2009.

\bibitem{citenet}
D.~J. de~Solla~Price, ``Networks of scientific papers,'' {\em Science},
  vol.~149, pp.~510--515, jul 1965.

\bibitem{flightnet}
C.~Li-Ping, W.~Ru, S.~Hang, X.~Xin-Ping, Z.~Jin-Song, L.~Wei, and C.~Xu,
  ``Structural properties of us flight network,'' {\em Chinese Physics
  Letters}, vol.~20, no.~8, p.~1393, 2003.

\bibitem{metnet1}
J.~Forster, ``Genome-scale reconstruction of the saccharomyces cerevisiae
  metabolic network,'' {\em Genome Research}, vol.~13, pp.~244--253, feb 2003.

\bibitem{metnet2}
J.~Triana, A.~Montagud, M.~Siurana, D.~Fuente, A.~Urchueguia, D.~Gamermann,
  J.~Torres, J.~Tena, P.~F. de~Cordoba, and J.~F. Urchueguia, ``{{G}eneration
  and {E}valuation of a {G}enome-{S}cale {M}etabolic {N}etwork {M}odel of
  {S}ynechococcus elongatus {P}{C}{C}7942},'' {\em Metabolites}, vol.~4,
  pp.~680--698, Aug 2014.

\bibitem{protnet1}
N.~Przulj, D.~Wigle, and I.~Jurisica, ``Functional topology in a network of
  protein interactions,'' {\em Bioinformatics}, vol.~20, pp.~340--348, feb
  2004.

\bibitem{protnet2}
F.~Hormozdiari, P.~Berenbrink, N.~Przulj, and S.~C. Sahinalp, ``Not all
  scale-free networks are born equal: The role of the seed graph in {PPI}
  network evolution,'' {\em {PLoS} Computational Biology}, vol.~3, no.~7,
  p.~e118, 2007.

\bibitem{gridnet1}
R.~Albert, I.~Albert, and G.~L. Nakarado, ``Structural vulnerability of the
  north american power grid,'' {\em Phys. Rev. E}, vol.~69, p.~025103, Feb
  2004.

\bibitem{gridnet2}
G.~A. Pagani and M.~Aiello, ``The power grid~as a complex network: A survey,''
  {\em Physica A: Statistical Mechanics and its Applications}, vol.~392,
  pp.~2688--2700, jun 2013.

\bibitem{ohno}
S.~Ohno, {\em Evolution by Gene Duplication}.
\newblock Springer, 2013.

\bibitem{wen}
W.-H. Li, {\em Molecular Evolution}.
\newblock Sinauer Associates, 1997.

\bibitem{ispolatov}
I.~Ispolatov, P.~L. Krapivsky, and A.~Yuryev, ``Duplication-divergence model of
  protein interaction network,'' {\em Phys. Rev. E}, vol.~71, p.~061911, Jun
  2005.

\bibitem{taylor}
J.~S. Taylor and J.~Raes, ``Duplication and divergence: The evolution of new
  genes and old ideas,'' {\em Annual Review of Genetics}, vol.~38, no.~1,
  pp.~615--643, 2004.
\newblock PMID: 15568988.

\bibitem{Ferreira_2016}
R.~M. Ferreira, R.~M. de~Almeida, and L.~G. Brunnet, ``Analytic solutions for
  links and triangles distributions in finite barabási–albert networks,''
  {\em Physica A: Statistical Mechanics and its Applications}, vol.~466,
  pp.~105 -- 110, 2017.

\bibitem{Ferreira_2003}
R.~M. Ferreira, J.~L. Rybarczyk-Filho, R.~J.~S. Dalmolin, M.~A.~A. Castro,
  J.~C.~F. Moreira, L.~G. Brunnet, and R.~M.~C. de~Almeida, ``Preferential
  duplication of intermodular hub genes: An evolutionary signature in
  eukaryotes genome networks,'' {\em PLOS ONE}, vol.~8, pp.~1--11, 02 2013.

\bibitem{Vazquez_2003}
A.~V\'azquez, ``Growing network with local rules: Preferential attachment,
  clustering hierarchy, and degree correlations,'' {\em Phys. Rev. E}, vol.~67,
  p.~056104, May 2003.

\bibitem{modeling}
A.~V{\'a}zquez, A.~Flammini, A.~Maritan, and A.~Vespignani, ``Modeling of
  protein interaction networks,'' {\em Complexus}, vol.~1, no.~1, pp.~38--44,
  2003.

\bibitem{Note1}
The bar over a term in the equation indicates that the average of the process
  will be considered.

\bibitem{euler}
L.~Euler, ``Solutio problematis ad geometriam situs pertinentis,'' {\em
  \it{Commentarii Academiae Scientiarum Imperialis Petropolitanae}}, vol.~8,
  pp.~128--140, 1736.

\bibitem{MacCluer2000}
C.~R. MacCluer, ``The many proofs and applications of perron's theorem,'' {\em
  {SIAM} Review}, vol.~42, pp.~487--498, jan 2000.

\end{thebibliography}
\end{document}